# Weak localization on moiré superlattice in twisted double bilayer graphene

Masaki Kashiwagi[1,a)], Toshihiro Taen[1], Kazuhito Uchida[1], Kenji Watanabe[2], Takashi Taniguchi[3], and Toshihito Osada[1,b)]

[1]*Institute for Solid State Physics, University of Tokyo, Kashiwa, Chiba 277-8581, Japan*

[2]*Reserch Center for Functional Materials, National Institute for Materials Science, Tsukuba, Ibaraki 305-0044, Japan*

[3]*International Center for Materials Nanoarchitectonics, National Institute for Materials Science, Tsukuba, Ibaraki 305-0044, Japan*

**Abstract**

Moiré superlattice created by twist stacking has multiple physical properties. These physical properties depend on the twist angle, hence investigation of the twist angle dependency is important for the deep understanding of physical phenomena in moiré superlattice. In this work, negative magnetoresistance owing to weak localization (WL) was investigated in twisted double bilayer graphene (TDBG) as a function of the twist angle. The ratio of the intervalley scattering time to the intravalley scattering time, estimated using the WL formula for bilayer graphene, tended to decrease as the twist angle increased. This feature is qualitatively explained by the enhancement of intervalley scattering due to the reduction of the intervalley distance in the moiré Brillouin zone (BZ) of the TDBG. This indicates that WL in the TDBG occurs for the moiré superlattice with the reconstructed BZ.

___

Authors to whom correspondence should be addressed: [a]kashiwagi@issp.u-tokyo.ac.jp, [b]osada@issp.u-tokyo.ac.jp

Van der Waals heterostructures[1] (VdWHs) of atomically thin films with rotational misalignment have a moiré pattern, which induces rich physical properties. Recent progress in the transfer techniques of atomically thin films enabled the fabrication of arbitrary VdWHs, in other words, various combinations of thin films and rotational misalignment[2]. Recent studies reported that twisted bilayer graphene (TBG), which consists of two monolayer graphene (MLG) stacked with a twist, shows a correlated insulating state[3,4], superconductivity[5,6], and ferromagnetism[7,8]. Similar to TBG, twisted double bilayer graphene (TDBG), which is a twisted stack of two bilayer graphene (BLG) films, also shows multiple physical properties[9,10,11]. These physical properties depend on the twist angle of TBG or TDBG, and the twist angle has attracted remarkable attention as a degree of freedom of the material.

Weak localization (WL) and weak antilocalization (WAL) effects are characteristic phenomena that reflect the scattering process in the system. WL is caused by the constructive interference between time-reversal scattering processes, forming a standing wave, and exhibits a negative magnetoresistance corresponding to the destruction of interference. In contrast, the WAL corresponds to the destructive interference at a zero magnetic field by the (pseudo-) spin rotation, and exhibits the positive magneto resistance. In MLG, WAL appears when intravalley scattering is dominant because of the pseudo-spin rotation as well as WL when intervalley scattering is dominant[12,13], whereas only WL is observed in BLG and multilayer graphene, where both intravalley and intervalley scattering contribute to WL[14,15,16]. As for TBG systems, the observation of WL suggested that intervalley scattering is enhanced in TBG compared to MLG[17,18]. Scattering by sharp point defects, local deformation, and bending in an artificially stacked graphene system were proposed as possible origins of this enhancement with no quantitative analysis, but the related mechanism has not been clarified yet. These works considered valleys in the original Brillouin zone (BZ) of each layer.

In this study, we investigated the twist angle dependence of negative magnetoresistance owing to WL in TDBG, and discussed intervalley scattering in the mini BZ of the moiré superlattice. The electronic structure and properties of twisted stacking systems depend on the twist angle caused by changes in the moiré pattern. In order to evaluate electron scattering in twisted stacked graphene, we chose TDBG for a more direct observation of WL than in TBG. The BLG has a parabolic band in the vicinity of the K(K') valley, and the Berry phase surrounding the quadratic band contact points at K(K') is equal to zero (or $2\pi$), leading to WL. However, in MLG with linear Dirac cone dispersion, the Berry phase is equal to $\pi$, resulting in WAL. In TBG and TDBG, the band contact points at the corners of the moiré BZ ($\overline{K}$ and $\overline{K}'$ points) originate from those at the K and K' points in the original BZ of each layer. Therefore, the TDBG is expected to exhibit WL without any mixture of WAL if the Berry phase surrounding the $\overline{K}$ ($\overline{K}'$) point is zero for a rather large twist angle.

TDBG samples were fabricated by the "tear and stack" technique[2]. Both graphene and hexagonal boron nitride (hBN) were mechanically exfoliated and transferred on a $SiO_2$/doped-Si substrate. The thickness of the flakes was evaluated by optical contrast under a microscope, which was confirmed by Raman spectroscopy in the case of graphene. The schematic structure of our device is shown in Fig. 1(a). The TDBG was fabricated using the following processes: First, the top hBN was picked up using a polydimethylsiloxane hemisphere coated with polymethylmethacrylate. Subsequently, a part of the BLG was contacted hBN and picked up with tearing the BLG film. The remainder of the BLG remained on the substrate. In sequence, the substrate was rotated and the hBN/BLG structure was released on the remaining BLG. The TDBG was capped by hBN with a thickness of approximately 10–20 nm and Au/Cr electrodes were deposited by electron beam evaporation. An optical microscopy image of the device is shown in Figure 1(b). The twist angles of the devices were $\theta$ = 0.93º, 2.93º, 4.84º, and 7.15º, respectively. We also prepared a BLG and a four-layer graphene(4LG) sample capped with hBN as a reference. The resistance was measured using the two-terminal method at several temperatures from 1.8 K to 30 K, and the carrier density $n$ was tuned to $1 \times 10^{12}$ cm$^{-2}$ for all samples by adjusting the back-gate voltage. This carrier density was determined from the amount of carriers injected with the charge neutral point(CNP) as the origin. The CNP was found as the peak in the back-gate voltage dependence of resistance(Figure 1(c)) From the carrier density and the twist angle, we calculated the filling factor $\nu = n/4n_\mathrm{M}$, which indicates the ratio of carrier occupation in the lowest energy band. Here $n_\mathrm{M} = 8\sin^2\frac{\theta}{2}/\sqrt{3}a^2$ is the density of moiré unit cell. As shown in Table I, the filling factor $\nu$ was less than a half in all TDBG samples. Therefore, the carriers occupied the lowest moiré band around the $\overline{K}$ and $\overline{K}'$ points.

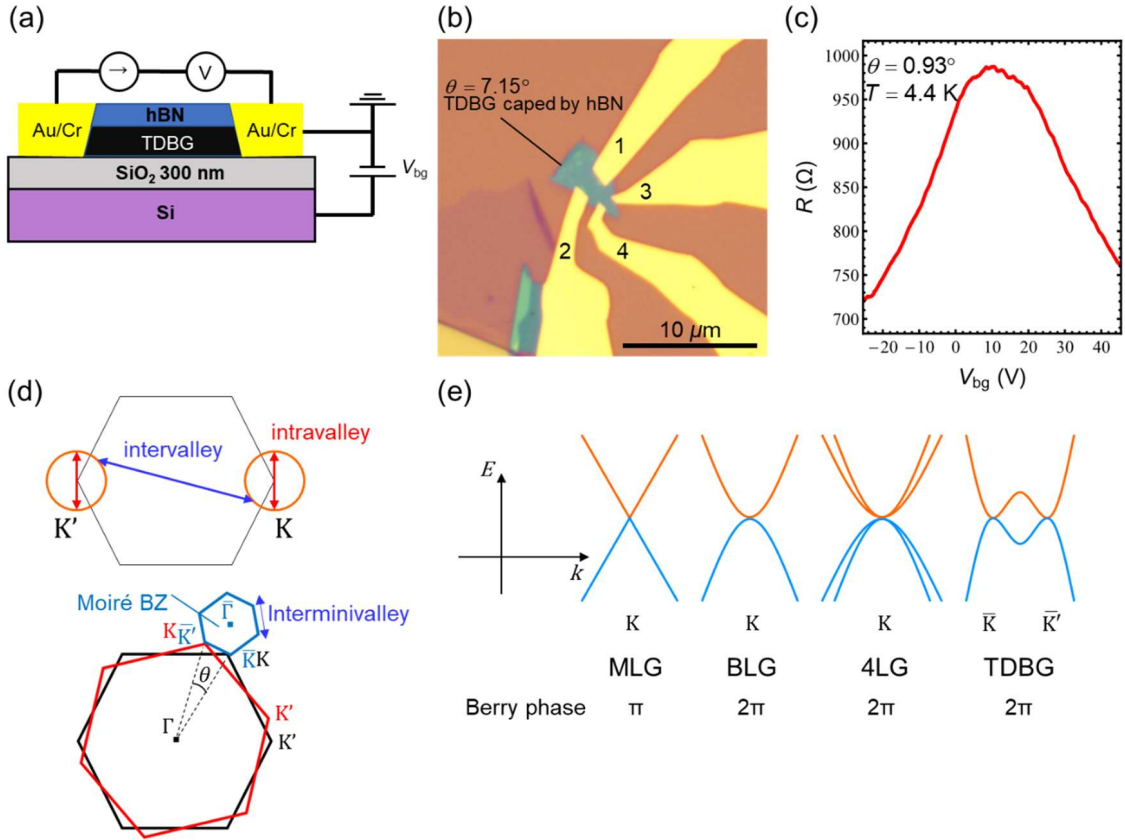

**FIG. 1.** (a) Schematic of the device structure. (b) Optical microscope image of TDBG device capped by hBN. In the case of this sample, the resistance between terminal 1 and 2 was measured, and back-gate voltage was induced through terminal 4. (c) Back-gate-voltage dependence of longitudinal resistance measured by two-terminal method at 4.4 K in the TDBG sample with $\theta = 0.93°$. (d) Schematics of the valley structure, the intervalley scattering and the intravalley scattering. Top figure shows the valley structure of MLG, BLG and 4LG. Bottom figure shows the valley structure of TDBG. (e) Schematics of the band structure of MLG, BLG, 4LG, and TDBG. Berry phase in the vicinity of K(K') point is shown respectively.

**Table I** Dephasing time $\tau_\phi$, intervalley scattering time $\tau_i$, and intravalley scattering time $\tau_*$ at T = 4.4 K of four TDBG samples with different twist angles, obtained from the fitting of Eq. (1). In addition, the BLG and 4LG data are presented for reference.

|  | TDBG | | | | BLG | 4LG |
|---|---|---|---|---|---|---|
| $\theta$ (°) | 0.93 | 2.93 | 4.84 | 7.15 | - | - |
| ΔK | 0.016 | 0.05 | 0.083 | 0.12 | 1 | 1 |
| Mobility (cm²/V·s) | 2.1×10³ | 8.3×10² | 6.9×10³ | 1.2×10³ | 4.6×10³ | 1.9×10³ |
| Carrier density (cm⁻²) | 7.5×10¹¹ | 7.5×10¹¹ | 7.5×10¹¹ | 7.5×10¹¹ | 7.5×10¹¹ | 7.5×10¹¹ |
| Filling factor | 0.38 | 0.038 | 0.014 | 0.0063 | - | - |
| $\tau_\phi$ (ps) | 3.9 | 3.7 | 4.2 | 4.6 | 4.6 | 1.7 |
| $\tau_i$ (ps) | 2 | 3.2 | 3.0 | 4.2 | 47 | 22 |
| $\tau_*$ (ps) | 0.24 | 0.2 | 0.14 | 0.099 | 0.4 | 0.3 |
| $\tau_*/\tau_i$ | 0.1 | 0.06 | 0.05 | 0.03 | 0.009 | 0.01 |

Figures 2(a), (b), and (c) show the magnetoconductivity $\Delta\sigma(B) = \sigma(B) - \sigma(0) = \frac{1}{\rho_{xx}(B)} - \frac{1}{\rho_{xx}(0)}$ of the TDBG with the twist angles of $\theta = 7.15°$, $\theta = 4.84°$, and $\theta = 2.93°$, respectively under normal magnetic fields. The magnetoconductivity increased as the magnetic field increased for all TDBG samples at low temperatures. This behavior is originated from the WL effect. The increase of magnetoconductivity became small as the temperature increased and negligible above $T$ = 30 K.

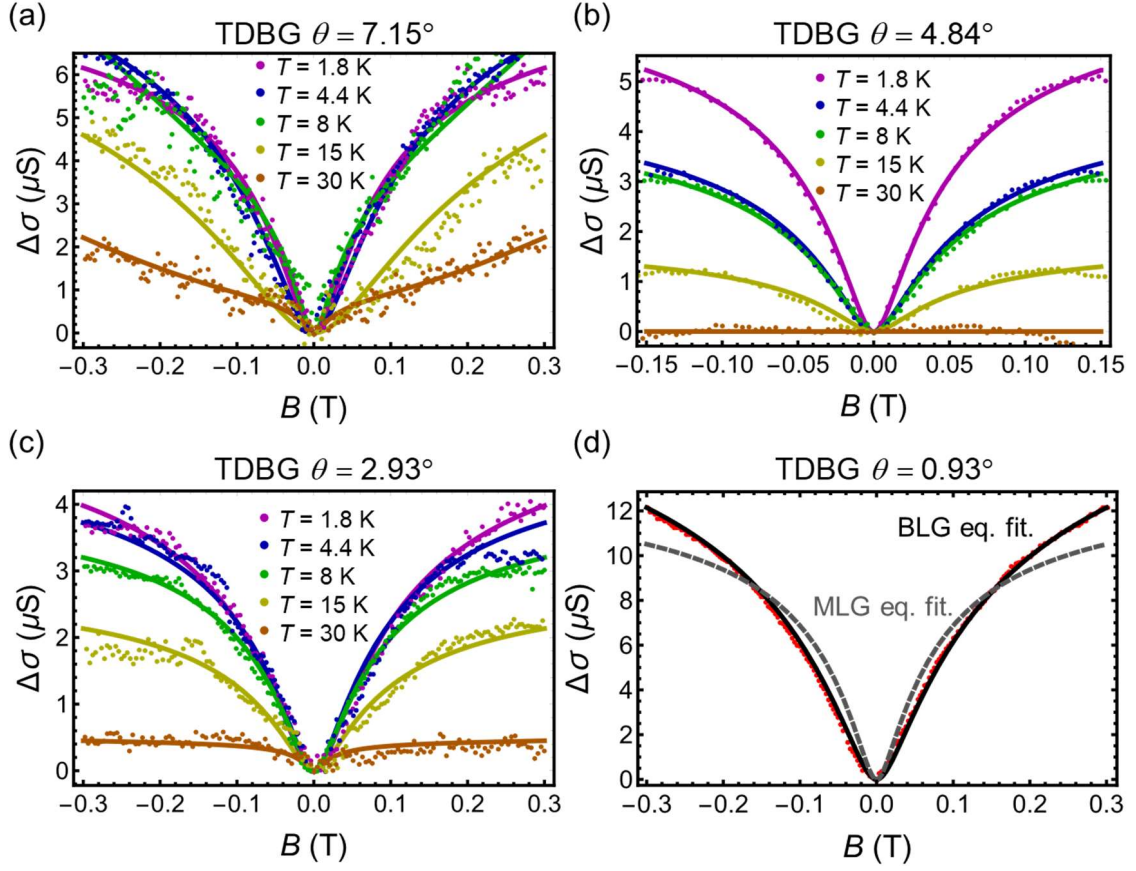

**FIG. 2.** Magnetic-field dependence of the conductivity of TDBG samples with (a) $\theta =$ 7.15º, (b) $\theta =$ 4.84º, and (c) $\theta =$ 2.93º at several temperatures. The solid lines indicate the fitting results of Eq. (1). (d) Comparison between the results using the fitting function for BLG (black solid line) and MLG (gray dashed line). The red dots are the data of TDBG with $\theta =$ 0.93º at $T =$ 4.4 K.

As indicated by the solid curves in Fig. 2, the observed positive magnetoconductivity was fitted using the following equation:[19,20,21]

$$\Delta\sigma(B) = \frac{e^2}{\pi h}\left\{F\left(\frac{B}{B_\phi}\right) - F\left(\frac{B}{B_\phi + 2B_i}\right) + 2F\left(\frac{B}{B_\phi + B_i + B_*}\right)\right\}. \quad (1)$$

Here $F(z) = \ln z + \Psi\left(\frac{1}{2} + \frac{1}{z}\right)$, where $\Psi$ is the digamma function. $B_{\phi,i,*} = \frac{\hbar}{4eD\tau_{\phi,i,*}}$, where $D$, $\tau_\phi$, $\tau_i$, and $\tau_*$ are the diffusion coefficient, dephasing time due to inelastic scattering, elastic intervalley scattering time, and elastic intravalley scattering time, respectively. This equation was originally proposed for the WL in AB-stacking BLGs, which has parabolic band dispersions in the vicinity of the K (K') points with the zero Berry phase surrounding them. According to the continuous model[22,23], the TDBG system also has parabolic dispersions around the $\overline{K}$ and $\overline{K}'$ points at the corners of the moiré BZ, which originate from the K (K') points of each BLG layer. According to recent theories[24,25,26], the Berry phase surrounding the $\overline{K}$ and $\overline{K}'$ points in the AB-AB stacking TDBG is zero, because the Berry curvature remains zero when applied perpendicular electric field is small. Therefore, the WAL contribution should be precluded, and we can safely use the WL formula of Eq. (1) as a fitting function.

To validate this choice of fitting function, we also performed fitting using the formula for MLG, in which the sign of the third term in Eq. (1) was reversed. This sign change is due to the WAL related to the Berry phase $\pi$ in the MLG. As shown in Fig. 2(d), Eq. (1) provided a better fit than the formula for the MLG.

The fitting parameters of each sample are listed in Table 1. There is a small increasing trend on $\tau_i$ and a decreasing trend on $\tau_*$ as the twist angle of TDBG increased. However, the absolute values of these parameters depend on the quality of samples, such as the amount of scattering impurities, defects, and dislocations present. Therefore, we employed the ratio $\tau_*/\tau_i$ such as to represent the enhancement of the intervalley scattering against intravalley scattering by compensating for sample dependence.

The experimental results indicated that $\tau_*/\tau_i$ in TDBG was larger than that in BLG. This suggests that intervalley scattering was enhanced in TDBG compared to that in BLG, and this enhancement is similar to that in the TBG reported in previous studies[17,18]. On the other hand, it was also shown that $\tau_*/\tau_i$ in TDBG decreased as the twist angle of TDBG increased. In previous studies[18], the enhancement of intervalley scattering was ascribed to extrinsic origins, that is, sharp point defects, local deformation, and bending by artificial stacking of two graphene layers. However, we propose an intrinsic origin based on the formation of a moiré BZ to explain the twist angle dependence of intervalley scattering in TDBG.

In TBG or TDBG with a moiré superlattice, the moiré BZ can be considered in the continuous model, as schematically shown in Fig. 3(a). The distance between the $\overline{K}$ and $\overline{K}'$ points of moiré BZ is given by $\Delta K = 2K_0/\sin(\theta/2)$. Therefore, the size of the moiré BZ depends on the twist angle and is smaller than that of the original BZ. Most electronic structures and properties can be described by the moiré BZ. The WL on the moiré superlattice can be discussed by considering intravalley and intervalley scattering in the moiré BZ with $\overline{\text{K}}$ and $\overline{\text{K}}'$ valleys. Here, intervalley scattering must be largely affected by the reduction of the distance between the $\overline{\text{K}}$ and $\overline{\text{K}}'$ valleys in the moiré BZ. When the Fourier component of the scattering potential $v(\mathbf{q})$ is a decreasing function of $q = |\mathbf{q}|$, intervalley scattering with $q = |\overline{\mathbf{K}} - \overline{\mathbf{K}}'| = \Delta K$ is considered to be enhanced at small twist angles.

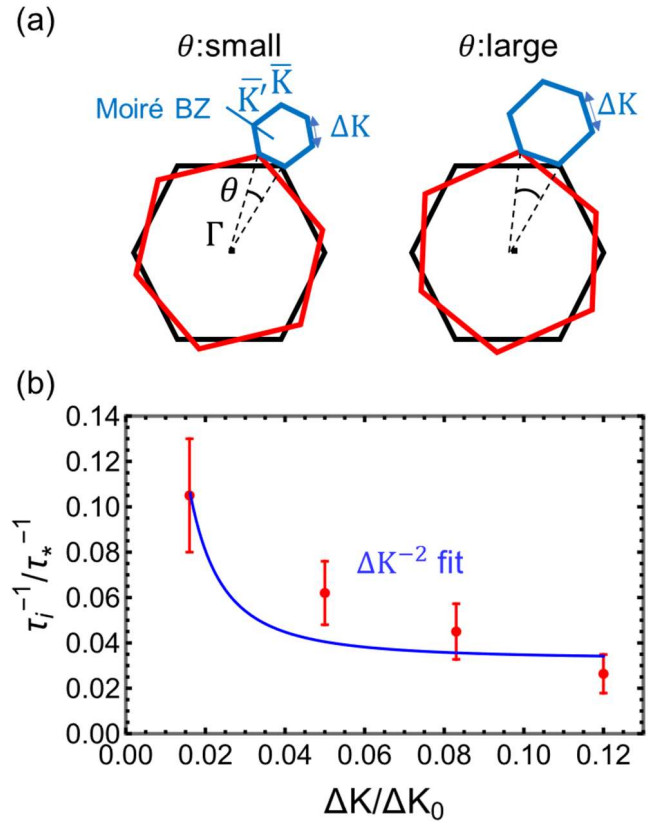

**FIG. 3.** (a) Schematics of moiré BZ of two TDBG with different twist angles. The edge length of moiré BZ, $\Delta K$, corresponds to the distance between **K** (or **K'**) of two BLG. (b) $\Delta K$ dependence of $\tau_*/\tau_i$ in TDBG samples at $T = 4.4$ K. $\Delta K$ is normalized by $\Delta K_0 = |\mathbf{K} - \mathbf{K}'|$ in BLG. The solid line shows the fitting result of $1/(\Delta K)^2$.

For a more quantitative evaluation, we considered the long-range Coulomb-type scattering potential with inverse of distance dependence (~ 1/r). When we perform two-dimensional Fourier transform for Coulomb type potential as,

$$v(\mathbf{k}) = \frac{1}{2\pi} \iint \frac{1}{|\mathbf{r}|} \exp\{-i\mathbf{k} \cdot \mathbf{r}\} d\mathbf{r}$$

The result becomes $v(k) = 1/k$. In this case, the scattering probability associated to the WL is proportional to $|v(\mathbf{q})|^2 \propto 1/q^2$, and we estimated the twist angle dependency of the intravalley scattering and intervalley scattering due to Coulomb-type potential. Since we fixed the carrier density, the size of the Fermi pockets at $\bar{K}$ and $\bar{K}'$ is unchanged with the twist angle, which means that the values of $q$ related to the intravalley scattering (time-reversal pair in each intravalley) are irrespective of the twist angles. The intravalley scattering probability is hence not affected by the twist angle. By contrast, the intervalley scattering probability is proportional to the twist angle through the relation $|v(\bar{\mathbf{K}} - \bar{\mathbf{K}}')|^2 \propto 1/(\Delta K)^2$. Therefore, we can expect that the ratio between intravalley and intervalley scattering times $\tau_*/\tau_i$ is proportional to $1/(\Delta K)^2$ in the TDBG. Figure 3(b) shows the $\Delta K$ dependence of $\tau_*/\tau_i$ of the TDBG with different twist angles. Here, $\Delta K$ is normalized by $\Delta K_0 = |\mathbf{K} - \mathbf{K}'|$, which is the distance between points K and K' in the BLG. The solid curve indicates the fitting result, assuming $\tau_*/\tau_i \propto 1/(\Delta K)^2$. The observed trend of $\tau_*/\tau_i$ with different twist angles is roughly reproduced by this relation. This agreement suggests

that the WL in TDBG originates from quantum interference owing to multiple scattering in the moiré superlattice scale disorder, but not that in the original BLG. For example, the twist angle disorder[27] is considered as the candidate of creating the intervalley scattering.

From the above discussion, it can be expected that 4LG shows the highest $\tau_*/\tau_i$ because 4LG is considered as TDBG with $\theta = 0°$ in a certain view. However, in our experiment, $\tau_*/\tau_i$ in 4LG was not larger than the other TDBGs. In the result of fitting to Eq. (1), it is considered that the intervalley scattering in 4LG is captured in the same valley structure of MLG or BLG and not of TDBG. Therefore, it does not seem appropriate to consider 4LG as TDBG with $\theta = 0°$, and it is suggested that the difference between the fitting result of TDBG and 4LG originates from the existence of moiré superlattice.

For the realization of the interference in the moiré BZ, the characteristic scattering length must be comparable to or larger than the scale of the moiré superlattice in a real system. To verify this, we compared the intervalley scattering length $L_i = v_F \tau_i$ with the size of the moiré unit cell $L = a_0/(2\sin(\frac{\theta}{2}))$. Here, $v_F = \frac{\hbar}{m^*}\sqrt{\frac{\pi n}{2}}$ is the Fermi velocity, $m^*$ is the effective mass of the carrier, $n$ is the carrier density, and $a_0$ is the the lattice constant of graphene. At a twist angle of $\theta \approx 1°$, the effective mass increased to $m^* \approx 0.3 m_e$[28], where $m_e$ is the electron rest mass. For the case of $\theta = 0.93°$, using this effective mass, the intervalley scattering length was estimated as $L_i$ = 84 nm. Because it is longer than the moiré period, $L$ = 15 nm, electrons can sense the moiré potential during the intervalley scattering process. For the other twist angle samples with no remarkable mass enhancement, we adopted the effective mass of the BLG, $m^* = 0.03 m_e$[29]. We found that the intervalley scattering length was longer than the moiré period in all case. This fact justifies the use of Eq. (1) as the fitting function. In addition, we compared the intravalley scattering length $L_*$ and size of the moiré unit cell in the same manner. We found that the intravalley scattering length was longer than the moiré period for the cases of $\theta > 1°$, except for the case of $\theta = 0.93°$, where the estimated intravalley scattering length $L_*$ = 9 nm was less than the moiré period. Although the large error in the fitting parameters may be responsible for this unfavorable result, most plausible reason is that the fitting function for BLG of Eq. (1) can be hardly applied to a TDBG with a twist angle of less than 1° because the distance between $\overline{K}$ and $\overline{K}'$ points becomes very small, and the hybridization between the two BLG layers complicates the band structure[25,26]. Therefore, for a small twist angle $\theta < 1°$, we need the other picture of the WL in a moiré superlattice system with a very large period.

We note that our result does not exclude the possibility that the enhancement of the intervalley scattering in twisted stacking graphene is a result of sharp point defects, local deformation, and bending, as proposed in previous studies[18], in which the valleys in the original BZ of each layer were considered; however, to explain the twist angle dependence of the enhancement of intervalley scattering, it seems appropriate to consider that the scattering occurs in the mini-moiré BZ rather than the original BZ of each layer.

Related to our result, interminivalley scattering was studied by measurement of high-temperature magnetooscillations in recent studies[30,31], and strong interminivalley scattering was observed[30]. According to the result, the intervalley and intravalley scattering times were of similar order in the small-angle TBG with $\theta = 1.65º$. This agrees with our result that the intervalley and intravalley scattering times become close to comparable values as the twist angle decreases.

In summary, we studied the twist angle dependence of the WL in TDBG. Intervalley and intravalley scattering times were obtained for several TDBG samples with different twist angles using the WL formula for the BLG. The ratio of the intervalley scattering time to the intravalley scattering time tended to decrease as the twist angle increased, which can be explained by the twist-angle dependence of the distance between $\overline{\mathrm{K}}$ and $\overline{\mathrm{K}}'$ in the moiré BZ. This implies that the WL in the TDBG occurs for the moiré superlattice with the reconstructed BZ.

**Acknowledgements**

This work was partly supported by JSPS KAKENHI, Grant Numbers JP19K14655, JP20H01860, and JP21K18594.